\documentclass[%
 reprint,
 superscriptaddress,
 amsmath,amssymb,
 aps,
prl,
]{revtex4-2}

\usepackage{color}
\usepackage{graphicx}
\usepackage{dcolumn}
\usepackage{bm}
\usepackage[squaren]{SIunits}

\begin{document}

\preprint{APS/123-QED}

\title{Generating 10 GW-class isolated zeptosecond x-ray bursts through bootstrapping two plasma-based accelerator stages}

\author{Qianyi Ma}
\affiliation{State Key Laboratory of Nuclear Physics and Technology, and Key Laboratory of HEDP of the Ministry of Education, CAPT, School of Physics, Peking University, Beijing 100871, China}
\author{Yuhui Xia}
\affiliation{State Key Laboratory of Nuclear Physics and Technology, and Key Laboratory of HEDP of the Ministry of Education, CAPT, School of Physics, Peking University, Beijing 100871, China}
\author{Zhenan Wang}
\affiliation{State Key Laboratory of Nuclear Physics and Technology, and Key Laboratory of HEDP of the Ministry of Education, CAPT, School of Physics, Peking University, Beijing 100871, China}
\author{Yuekai Chen}
\affiliation{State Key Laboratory of Nuclear Physics and Technology, and Key Laboratory of HEDP of the Ministry of Education, CAPT, School of Physics, Peking University, Beijing 100871, China}
\author{Letian Liu}
\affiliation{State Key Laboratory of Nuclear Physics and Technology, and Key Laboratory of HEDP of the Ministry of Education, CAPT, School of Physics, Peking University, Beijing 100871, China}
\author{Zhiyan Yang}
\affiliation{State Key Laboratory of Nuclear Physics and Technology, and Key Laboratory of HEDP of the Ministry of Education, CAPT, School of Physics, Peking University, Beijing 100871, China}
\author{Chao Feng}
\affiliation{Shanghai Advanced Research Institute, Chinese Academy of Sciences, Shanghai 201210, China}
\author{Xinlu Xu}
\email[]{xuxinlu@pku.edu.cn}
\affiliation{State Key Laboratory of Nuclear Physics and Technology, and Key Laboratory of HEDP of the Ministry of Education, CAPT, School of Physics, Peking University, Beijing 100871, China}
\affiliation{Beijing Laser Acceleration Innovation Center, Huairou, Beijing, 101400, China}
\author{Xueqing Yan}
\email[]{x.yan@pku.edu.cn}
\affiliation{State Key Laboratory of Nuclear Physics and Technology, and Key Laboratory of HEDP of the Ministry of Education, CAPT, School of Physics, Peking University, Beijing 100871, China}
\affiliation{Beijing Laser Acceleration Innovation Center, Huairou, Beijing, 101400, China}
\affiliation{Institute of Guangdong Laser Plasma Technology, Baiyun, Guangzhou, 510540, China}
\author{Chan Joshi}
\affiliation{Department of Electrical Engineering, University of California Los Angeles, Los Angeles, California 90095, USA}
\author{Warren B. Mori}
\affiliation{Department of Electrical Engineering, University of California Los Angeles, Los Angeles, California 90095, USA}
\affiliation{Department of Physics and Astronomy, University of California Los Angeles, Los Angeles, California 90095, USA}

\date{\today}

\begin{abstract}
A method is proposed to generate coherent, intense zeptosecond x-ray pulses by using the electron beam produced by a laser wakefield accelerator (LWFA) stage as the driver for a beam driven plasma wakefield accelerator (PWFA) stage. The LWFA injector stage requires a readily available 100 TW laser driver to produce a GeV class self-injected electron beam. This beam is focused and used as the driver in a PWFA stage that utilizes a solid-density plasma with a modulated density downramp. The concept is shown to be capable of producing an ultra-short electron beam with unprecedented density ($10^{26}~\mathrm{cm^{-3}}$) and brightness ($ 10^{24}~\ampere/\meter^2/\radian ^2$), that is also pre-bunched  on 0.1 Angstrom scales. By colliding this pre-bunched extreme beam with an optical undulator, an intense zeptosecond pulse can be emitted if the spatially focusing region is matched with the lasing region of the beam. Multi-dimensional particle-in-cell simulations are performed of the entire concept to demonstrate that an isolated 10 GW-class zeptosecond pulse with a full-width-half-maximum duration of 700 zs can be generated in a tapered laser pulse. Such an intense zeptosecond pulse generation scheme may provide an essential probe for nuclear physics and quantum electrodynamics processes. 
\end{abstract}

\maketitle

Attosecond pulses have revolutionized ultrafast science by capturing and manipulating electron dynamics inside atoms, molecules and solids \cite{as-review,as-science,as-electron-AM-decay,as-photon-emission-delay}. The next frontier, zeptosecond ($1~\zepto\second=10^{-21}~\second$) pulses, represents the characteristic timescale of some phenomena in quantum electrodynamics (QED), and nuclear physics \cite{zs-application}. Such pulses would enable probing of electron-positron pair production and annihilation in vacuum \cite{zs-ep-pair}, birth time delay in high-field photoionization \cite{zs-birth-delay} and onset of compound nuclear reactions \cite{book:nuclear,Bohr-nuclear-model}. High-photon-energy zeptosecond pulses may lead to different forms of nuclear excitation \cite{zs-nuclear-excitation}, which facilitates the validation of corresponding nuclear physics theory. 

High-order harmonic generation (HHG) in gases \cite{as-science,gas-hhg-review}, which can routinely generate attosecond pulses may be extended to produce zeptosecond pulse trains based on the interference of emitted x-rays, albeit with an extremely low intensity \cite{zs-He} or using muonic atoms \cite{zs-muonic}. HHG in solid-density plasma \cite{lambda3,plasma-hhg-review} may produce zeptosecond pulse trains as shown by one-dimensional (1D) simulations \cite{zs-rom,zs-cascade}, however, their intensity is significantly degraded by high-dimensional effects \cite{hhg-3d,hhg-3d2,hhg-diffraction-limited} and the finite width of the vacuum–plasma interface \cite{hhg-decrease-efficiency,hhg-density-gradient,hhg-scale-length,hhg-optimal-scale-length}. Generation of high-power zeptosecond pulses remains one of the grand challenges in modern physics.

Tens of fs intense laser pulses \cite{lwfa-1979} or charged particle beams \cite{pwfa-1985} can produce high-quality electron beams with sub-fs duration through plasma-based acceleration (PBA) \cite{joshi2020perspectives}. These short and dense electron beams offer new opportunities for high-power attosecond radiation generation \cite{xu2021generation,pwfa-compression-tw-as,self-selection,malaca2024coherence,Peng-PhysRevLett.131.145003}. The $10\sim 100~\mega\electronvolt/\micro\meter$ energy chirps that are often present in PBA beams, allow them to be compressed to $\sim100$ as, which enables the emission of TWs of soft x-ray radiation in a magnetic undulator with similar durations \cite{pwfa-compression-tw-as}. Moreover, a counter-propagating laser pulse can select a 100-as fraction of the electrons to lase if the chirped beam is pre-bunched \cite{self-selection}. Recent works propose using the superluminal high-density tail of the nonlinear plasma wake in a density upramp to generate subcycle ultraviolet radiation bursts \cite{malaca2024coherence, Peng-PhysRevLett.131.145003}. However, a further reduction of two to three orders of magnitude in pulse duration \cite{lwfa-fel-sase, lwfa-fel-seeded, vh62-gz1p} is needed to reach the zeptosecond realm.

In this Letter, we propose serially linking a laser wakefield accelerator (LWFA) and a beam driven wakefield accelerator (PWFA) to produce intense zeptosecond hard x-ray pulses. Here, the electron beam generated from a LWFA stage using a 100 TW laser driver and a gas jet plasma is focused into a PWFA stage with a solid-density plasma that has been fabricated with a combination of a density downramp and density modulation. The e-beam generated in the LWFA stage has a current in excess of 50 kA and a duration shorter than 1 fs such that it can effectively drive downramp injection in a solid-density plasma to produce a $\sim$10 as electron beam, modulated on 0.1 Angstrom scales, with extremely high density ($\sim10^{26}~\centi\meter^{-3}$) and brightness ($\sim10^{24}~\ampere/\meter^2/\radian ^2$). The extremely dense and bright electron beams are of general interest in their own right in areas ranging from high energy physics \cite{PhysRevLett.122.190404}, high-energy-density physics \cite{xu2021generation}, strong-field QED \cite{PhysRevLett.122.190404, matheron2023probing} and ultra-bright $\gamma$-ray sources \cite{benedetti2018giant, PhysRevLett.126.064801, PhysRevResearch.6.L032013, 10.1063/5.0174508}. 

\begin{figure}[htbp]
\includegraphics[width=\linewidth]{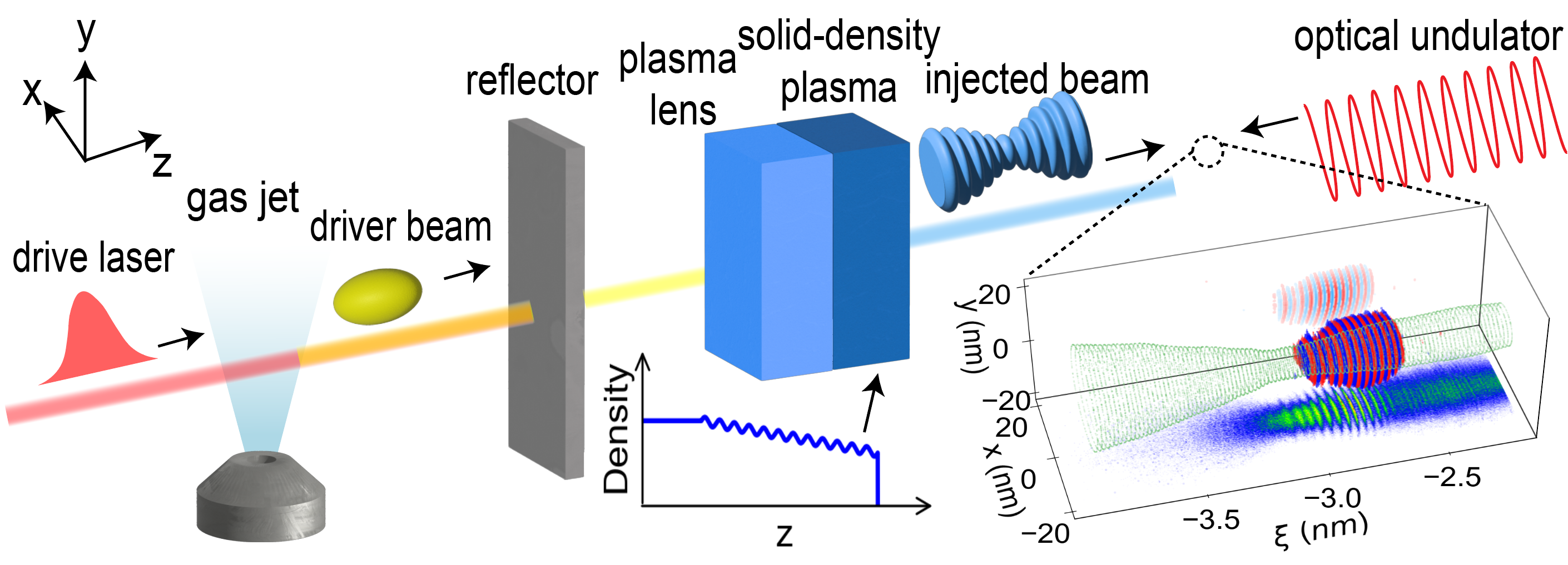}
\caption{\label{fig:1} Schematic of intense zeptosecond pulses generation (not to scale). The middle inset shows the tailored density of the solid-density plasma. The right inset illustrates the electric field of the zeptosecond pulse (red-blue) and the beam envelope (green) in the optical undulator, and the beam charge density projection (bottom) and the field in the central slice side (side).}
\end{figure}

The concept is illustrated in Fig. \ref{fig:1}, where the output e-beam is collided with an optical undulator where the prebunched beams can radiate superradiantly \cite{gover2019superradiant}. By controlling the overlap in space and time between the optical undulator and the lasing electrons, a radiation pulse with a duration substantially shorter than the beam duration can be generated. Multi-dimensional particle-in-cell (PIC) simulations demonstrate that it is feasible to generate an isolated zeptosecond pulse with $10^6$ hard x-ray photons. The simulations are performed using the fully relativistic PIC code OSIRIS \cite{osiris} with an advanced Maxwell solver \cite{xu2013numerical,xu-solver}; and the setup can be found in the Supplemental Material. Such a source may provide an essential tool for probing the dynamics of nuclear physics and QED processes. 

\begin{figure}[htbp]
\includegraphics[width=\linewidth]{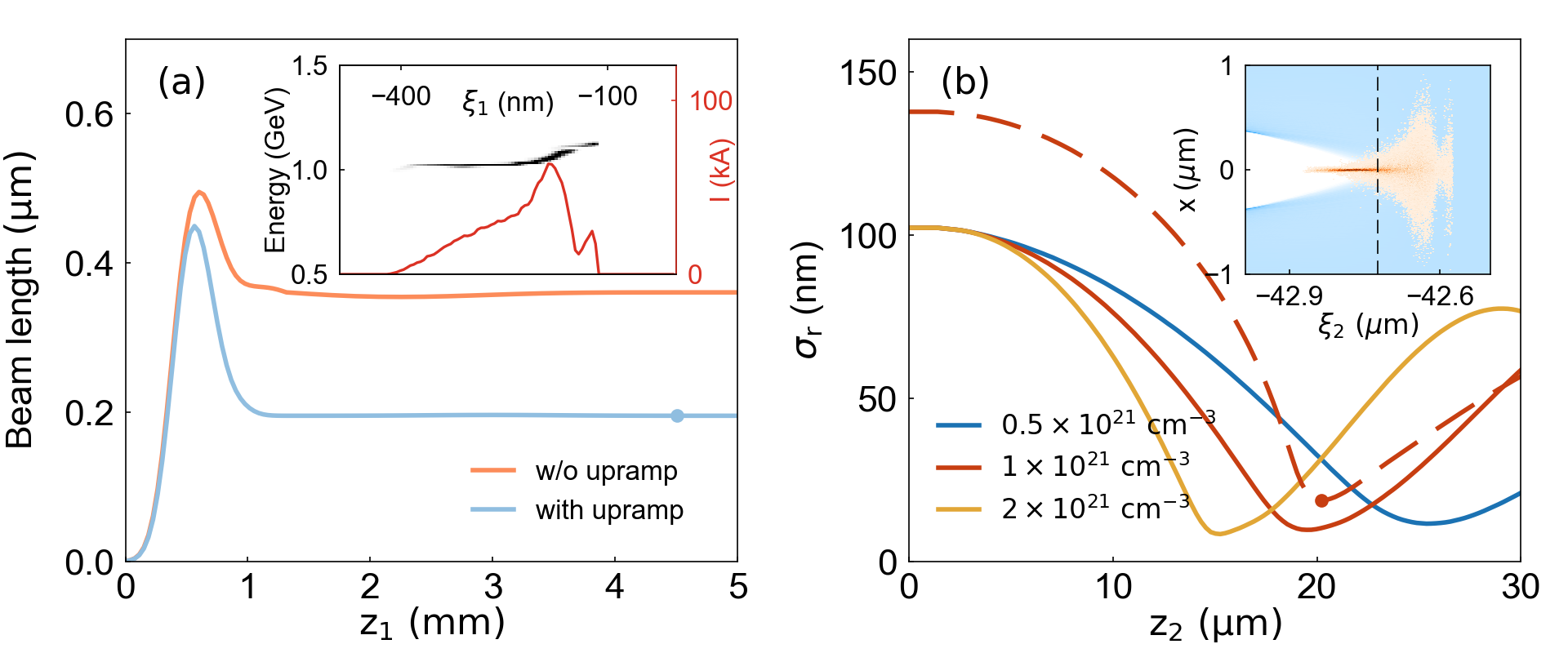}
\caption{\label{fig:2} Short and narrow GeV e-beam driver generation. (a) The evolution of the injected beam duration with (blue) and without (orange) an upramp. The inset shows the beam longitudinal phase space and the current profile at $z=4.5~\milli\meter$ when applying an upramp. (b) The beam spot size evolution in a thick plasma lens with different densities (solid lines for the second half of the beam and dashed line for the whole beam). The inset shows the beam density distribution (red) and the plasma (blue) at $z_2=20~\micro\meter$ in $1\times 10^{21}~\mathrm{cm^{-3}}$ case. }
\end{figure}

To understand the proposed concept it is useful to work backwards from the requirements for the second to the first stage. A high-current, high energy (GeV), e-beam with sub-fs duration and 10s of $\nano\meter$ spot size is required to excite a fully nonlinear plasma wake in a solid-density plasma ($\sim10^{23}~\centi\meter^{-3}$). Such a beam has not been produced using either PBAs or conventional accelerators. As shown in Fig. \ref{fig:1}, we propose relying on self-injection from LWFA in the nonlinear blowout regime to produce such a beam. Based on previous work \cite{PhysRevAccelBeams.26.111302}, we consider a 48 fs and 136 TW Ti:Sapphire drive laser pulse with a spot size of 11 $\micro\meter$ incident on an underdense plasma ($n_\mathrm{p1}=2\times 10^{18}~\centi\meter^{-3}$) (e.g., a gas jet). The laser is self-focused in the first 0.6 mm of the plasma, resulting in an expansion of the wake and injection of an e-beam with a length of 500 nm \cite{PhysRevAccelBeams.26.111302}. As the injected electrons are accelerated at the wake tail, the length of the wake gradually decreases due to the diffraction of the laser, causing some electrons to slip into the next wavelength (etch away) where they are defocused and lost. As shown by the red line in Fig. \ref{fig:2}(a), this causes the beam length to decrease to 400 nm at $z_1=1~\milli\meter$ and then remain unchanged, because the accelerating beam then advances faster than the wake tail and outruns it. In addition, if the plasma has a slight positive density gradient, then the etching process can be enhanced. As shown by the blue line in Fig. \ref{fig:2}(a), the bunch length is further reduced to 200 $\nano\meter$ when a density gradient of $5.3\times 10^{16}~\centi\meter^{-3}/\milli\meter$ is used. The injected electrons are eventually accelerated to 1.1 GeV energy with a peak current of 63 kA, normalized emittance of $\epsilon_\mathrm{N}\approx 0.3~\micro\meter$, and a small energy spread at $z_1=4.5~\milli\meter$ as shown in the inset of Fig. \ref{fig:2}(a). The normalized emittance is defined as $\epsilon_{\mathrm{N}}\equiv \sqrt{\langle x^2 \rangle \langle p_x^2 \rangle - \langle x p_x \rangle^2}$, which is the area of the beam transverse phase space, where $\langle \rangle$ represents an ensemble average over the beam particles.

The resulting 200-nm-long (sub-fs) e-beam has a Gaussian fitted spot size of $\sigma_{r_\mathrm{d1}}\approx 140~\nano\meter$ and a peak density of $n_\mathrm{d1}\approx 0.9\times 10^{22}~\centi\meter^{-3}$. The beam then passes through a 1-$\micro\meter$-thick metallic reflector which blocks the drive laser whereupon it enters an underdense thick plasma lens section with a density of $n_\mathrm{p2}=10^{21}~\centi\meter^{-3}$. The beam satisfies $\frac{\omega_\mathrm{p2}}{c}\sigma_{r_\mathrm{d1}}\approx 0.8$ and $n_\mathrm{d1}\approx 9n_\mathrm{p2}$, thus the head of the beam can excite a nonlinear plasma wake, whose linear focusing force focuses the second half of the beam [inset of Fig. \ref{fig:2}(b)] \cite{beam-focus}. For an unmatched beam with $\frac{\beta_\mathrm{d2}}{\beta_\mathrm{d1}} \ll 1$ ($\beta_{\mathrm{dj}}\equiv \sqrt{2\gamma_\mathrm{d}}\frac{c}{\omega_{\mathrm{pj}}}, \mathrm{j}=1,2$), the minimum spot size and its focal distance when neglecting the space-charge effects can be estimated as $\sigma_{r_\mathrm{d}}=\sigma_{r_\mathrm{d1}}\sqrt{\frac{n_\mathrm{p1}}{n_\mathrm{p2}}}\approx 6.3~\nano\meter$ and $z_\mathrm{m}=\frac{\pi\sqrt{2\gamma_\mathrm{d}}}{2}\frac{c}{\omega_\mathrm{p2}}\approx 17~\micro\meter$ (see Supplemental Material), where $\gamma_\mathrm{d}$ is the relativistic factor of the beam and $\omega_\mathrm{pj}=\sqrt{\frac{n_\mathrm{pj}e^2}{m\epsilon_0}}$ \cite{EMconstants}. The red solid and dashed lines in Fig. \ref{fig:2}(b) indicate the evolution of the spot sizes of the second half and the entire beam, respectively. These spot sizes reach minimum values (waists) of $10~\nano\meter$ and $19~\nano\meter$ respectively at $z_\mathrm{m} \approx 20~\micro\meter$. The beam waists are larger than the formulas due to the space-charge forces between electrons. The inset in Fig. \ref{fig:2}(b) illustrates the transverse density distribution of the beam at the focal distance, showing a tightly focused second half of the beam and a more diffuse first half (see Supplemental Material). This nonuniform focusing decreases the effective length of the beam when driving a PBA in the higher density plasma in the final stage. After focusing, the beam peak density increases by two orders of magnitude, reaching $\sim 10^{24}~\centi\meter^{-3}$. We also present the beam spot size evolution for plasma lenses with other densities in Fig. \ref{fig:2}(b), which shows the waist of the second half beam is smaller (larger) and the focal distance is shorter (longer) when $n_\mathrm{p2}$ is higher (lower). The conclusion is that sub-fs beams can be focused to $\sim10$ nm over a wide range of $n_\mathrm{p2}$.

In the second half of the second stage, the resulting dense and compact electron beam is used to self-inject and accelerate an electron beam with unprecedented parameters. We choose density downramp as the injection scheme \cite{PhysRevE.58.R5257, PhysRevLett.86.1011} since the injected beam properties can be scaled when varying the background plasma density $n_\mathrm{p}$ \cite{downramp}, e.g., the peak density $n_\mathrm{b}\propto n_\mathrm{p}$, the brightness $B_\mathrm{5D}\equiv \frac{2I}{\epsilon_\mathrm{N}^2}\propto n_\mathrm{p}$ and the 3D sizes $\sigma_{x,y,z} \propto n_\mathrm{p}^{-1/2}$. Previous work shows the density of the beam injected in a gas-density plasma is 3 orders of magnitude higher than the plasma density \cite{downramp}. We choose $n_\mathrm{p3}=6\times 10^{22}~\centi\meter^{-3}$ to achieve $\frac{\omega_\mathrm{p3}}{c}\sigma_{r_\mathrm{d}}\approx 0.5$. A density downramp with a length of $L=150\frac{c}{\omega_\mathrm{p3}}$ and a normalized gradient of $g\equiv \frac{\Delta n_\mathrm{p3}}{\omega_\mathrm{p3}L/c}= 5\times 10^{-4}$ is utilized. As shown in Fig. 3(a), a 3-nm-long electron beam with $n_\mathrm{b} \approx 10^{26}~\centi\meter^{-3}$ is injected at the end of the downramp and it has an energy chirp of $50~\mega\electronvolt/\nano\meter$ due to the longitudinal mapping between the initial positions of the electrons and their relative positions after injection $\xi\equiv z -c t$ \cite{downramp}. 

If a small-amplitude sinusoidal density modulation ($10^{-3}n_\mathrm{p3}$) with a period of $\lambda_\mathrm{m}=22~\nano\meter$ is superimposed on the downramp, a pre-bunched structure is formed along the injected beam as shown by modulations to the current profile [red line in Fig. 3(a)]. This is due to the injection being turned on and off at the plasma density modulation period \cite{xu2022generation}. A sinusoidal modulation profile is used for simplicity here, while other types of modulation profiles should work equally well. This density-modulated solid target can be realized by micro-nano fabrication techniques, such as physical vapour deposition \cite{deposition,target} and molecular beam epitaxy \cite{herman2012molecular}. Previous theoretical and simulation results show the injected electron beam is compressed by a factor $h_\mathrm{c}=2\gamma_\mathrm{w}^2\approx \frac{2}{k_\mathrm{p3}\lambda_\mathrm{w}g}$ \cite{downramp}, where $\lambda_\mathrm{w}$ and $\gamma_\mathrm{w}$ are the wake wavelength and the relativistic factor of the wake tail, respectively. For the parameter used in the simulation, the measured $h_\mathrm{c}\approx 680$ compared to the predicted value of 610. The resulting bunching factor $b(\lambda)\equiv \frac{1}{N}{\lvert \sum_{j=1}^N \mathrm{exp}(i\frac{2\pi c}{\lambda}\xi_j)\rvert}$, which quantifies the modulation, has a peak value $b\approx0.018$ at $\lambda \approx 0.32~\angstrom$, where $N$ is the electron number. The parameters of the beam, i.e., the energy chirp and the pre-bunching wavelength, can be adjusted through $n_\mathrm{p3}, g$ and $\lambda_\mathrm{m}$ \cite{xu2022generation}. We emphasize that the normalized emittance of the injected beam is $\epsilon_\mathrm{N}\approx 0.2~\nano\meter$, which in combination with its $20~\kilo\ampere$ current yields a brightness of $B_\mathrm{5D}\approx 10^{24}~\ampere/\meter^2/\radian^2$.

\begin{figure}
\includegraphics[width=\linewidth]{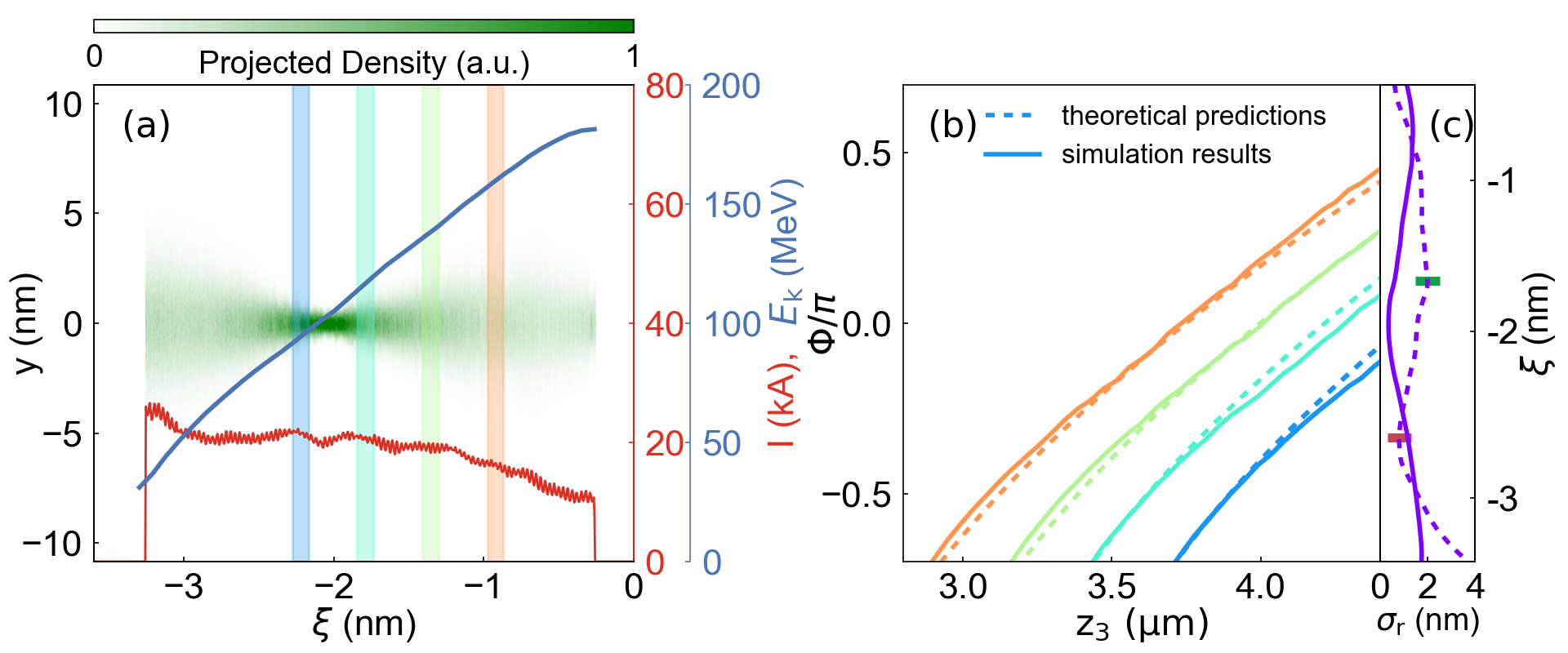}
\caption{\label{fig:3} Properties of the extreme e-beam produced in a solid-density PBA. (a) The projected charge density distribution (green), the current (red) and the energy profile (blue) of the injected beam at the end of the downramp. (b) The betatron phase evolution of 4 slices marked in (a). The solid lines are the simulation results and the dashed lines are the theoretical predictions. (c) The beam spot size at the end of the downramp (solid line) and after a drift of 0.6 $\micro\meter$ (dashed line).}
\end{figure}

In downramp injection, the electrons are not initially transversely matched with the linear focusing force of the plasma wake, causing them to rotate in the transverse phase space, i.e., to execute beam envelope betatron oscillations. The longitudinal mapping between $z_i$ and $\xi$ ensures that the electrons injected at different times reside at different axial positions \cite{downramp}. Thus, each axial slice has its own betatron phase given by $\Phi(z,\xi)=\Phi_\mathrm{i} + \int_{z_\mathrm{i}(\xi)}^z \mathrm{d}z' \frac{\omega_\mathrm{p3}}{c\sqrt{2\gamma}}$, where $\Phi_\mathrm{i}$ is the initial phase and $z_\mathrm{i}(\xi)$ represents the injection position of a slice which grows from the beam head to tail. The evolution of $\Phi$ for different slices is shown in Fig. \ref{fig:3}(b), where good agreement is achieved between the simulation results (solid lines) and the theoretical predictions (dashed lines). Since the spot size converges when $\Phi \in (\frac{\pi}{2},\pi)\cup(\frac{3\pi}{2}, 2\pi)$, and diverges when $\Phi \in (0, \frac{\pi}{2}) \cup(\pi,\frac{3\pi}{2})$, the axially varying $\Phi$ manifests spatially as a periodic oscillation of the beam spot size $\sigma_r$ in $\xi$ with a period of $L_\mathrm{s}\approx \frac{\sqrt{2\gamma}\pi}{h_\mathrm{c}}\frac{c}{\omega_\mathrm{p3}}$ (see Supplemental Material). The solid line in Fig. \ref{fig:3}(c) shows that $\sigma_r$ varies with a period of $\sim 2.5~\nano\meter$ and its minimum value is 0.5 nm at $\Phi=0~(\xi=-2.0~\nano\meter)$. As the beam drifts in free space, the profile of $\sigma_r$ shifts. The dashed line shows $\sigma_r$ after a drift of 0.6 $\micro\meter$ where the minimum value is 0.9 nm at $\xi=-2.7~\nano\meter$ and -0.3 nm. 

Using OSIRIS simulations, we also collide this extremely dense, bright, and prebunched beam with an 800 nm linearly polarized laser pulse with a flattop envelope at 0.6 $\micro\meter$ away from the plasma exit. As the electrons oscillate inside the laser pulse, they emit forward radiation with a Doppler shifted wavelength of $\lambda_\mathrm{r} = \frac{\lambda_\mathrm{L}}{4\gamma^2}(1+\frac{a_\mathrm{L}^2}{2})$, where $\lambda_\mathrm{L}$ and $a_\mathrm{L}$ are the wavelength and the normalized vector potential of the laser, respectively. If $\lambda_\mathrm{r}$ is equal to the pre-bunched wavelength, the electrons emit radiation superradiantly which in turn can enhance the modulation. A seeded free-electron laser (FEL) instability occurs and produces high power radiation \cite{xfel-review}. The growth rate of the FEL instability is quantified by the Pierce parameter $\rho$ \cite{fullexpressionofrho} which scales with the beam density and the radiation wavelength as $\rho \propto n_\mathrm{b}^{1/3}\lambda_\mathrm{r}^{2/3}$. To achieve zeptosecond pulses the wavelength of the radiation must be $\lesssim 0.1~\angstrom$, thus electron beams with extremely high density are necessary to produce high power radiation before the beams diffract significantly. 

\begin{figure}
\includegraphics[width=\linewidth]{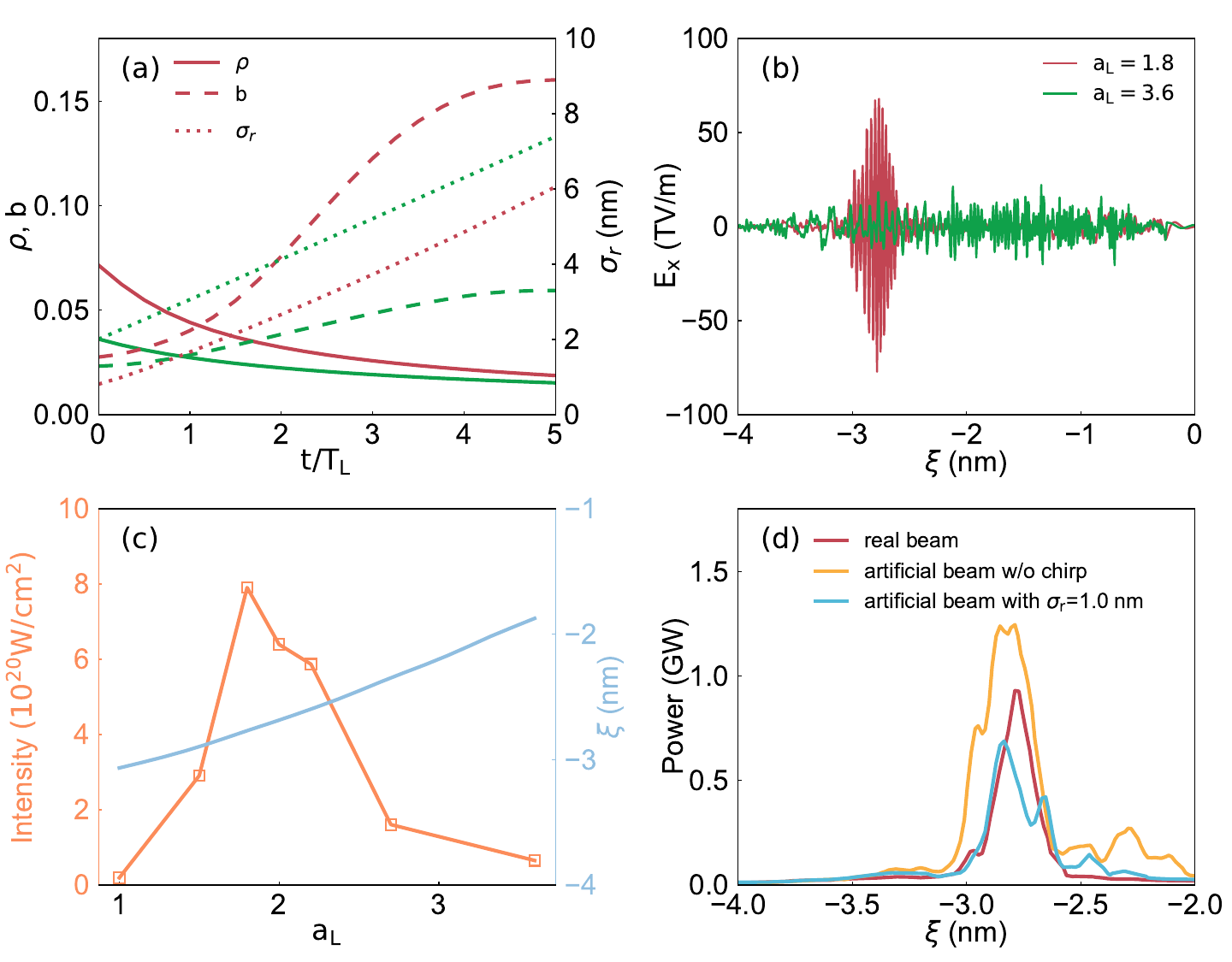}
\caption{\label{fig:4} Generation of zeptosecond pulses. (a) The evolution of the Pierce parameter $\rho$, the peak bunching factor $b$ and the spot size $\sigma_r$ of the electrons with resonant energy in $a_\mathrm{L}=1.8$ (red lines) and $a_\mathrm{L}=3.6$ (green lines) cases. (b) The radiation field $E_x$ at $t=4.5T_\mathrm{L}$ when $a_\mathrm{L}=1.8$ (red line) and $a_\mathrm{L}=3.6$ (green line). (c) The peak intensity of the radiation pulse (orange line) and the position of the resonant electrons (blue line) when varying $a_\mathrm{L}$. (d) The power profile of the real beam and two artificial beams at $t=4.5T_\mathrm{L}$.}
\end{figure}

Since the prebunched beam is chirped ($50~\mega\electronvolt/\nano\meter$), only a finite axial part can be resonant with a particular $a_\mathrm{L}$. We show two examples from simulations with $a_\mathrm{L}=1.8$ and 3.6 respectively, where the former is resonant with 67 MeV electrons which have a minimum $\sigma_r\approx 0.9~\nano\meter$ [red mark in Fig. \ref{fig:3}(c)] and the latter is resonant with 114 MeV electrons with $\sigma_r=2.0$ nm [green mark]. As the beam propagates against the laser, the beam evolves. A comparison of the evolution of $\sigma_r, \rho$ and the peak $b$ are shown in Fig. \ref{fig:4}(a) for both cases. When $a_\mathrm{L}=1.8$, the initially small size results in a ``large" $\rho \approx 0.08$, which is twice that for the $a_\mathrm{L}=3.6$ case. The beam expands rapidly as it propagates inside the laser because of its angular divergence and  space-charge effects, resulting in a growth of $\sigma_r$ and a decrease of $\rho$ for both cases. When $a_\mathrm{L}=1.8$, $b$ grows by one order of magnitude within 5 laser periods, whereas at $a_\mathrm{L}=3.6$, it merely doubles. The on-axis electric field $E_x$ of the radiation at $t=4.5T_\mathrm{L}$ are shown in Fig. \ref{fig:4}(b) for both cases, where $T_\mathrm{L}=400~ \nano\meter/c=1.33~\femto\second$ is the time it takes a beam particle to pass a laser wavelength. An isolated radiation pulse with an amplitude of $60~\tera\volt/\meter$ and a full-width-half-maximum (FWHM) duration of $550~\zepto\second$ is produced when $a_\mathrm{L}=1.8$. In contrast, when $a_\mathrm{L}=3.6$, the radiation develops over a large axial ($\xi$) region of the beam, and the field amplitude is lower by one order of magnitude. This comparison indicates it is critical to overlap the lasing region with the dense region of the e-beam for the intense zeptosecond pulse generation.

A detailed scan of the output pulse depending on the resonant region of the e-beam is performed by adjusting $a_\mathrm{L}$, and the peak intensity of the radiation pulse is presented in Fig. \ref{fig:4}(c). As $a_\mathrm{L}$ increases from 1.0 to 3.6, the resonant energy increases and the lasing region sweeps from $\xi=-3.0~\nano\meter$ to $\xi=-1.8~\nano\meter$. The radiation intensity reaches its maximum when $a_\mathrm{L}=1.8$ and decreases as $a_\mathrm{L}$ either increases or decreases. The FWHM of the $a_\mathrm{L}$-intensity line is $\Delta a_\mathrm{L}\approx 1$, which indicates we can tolerate a large variation of $a_\mathrm{L}$.  

Two prescribed beams are introduced to study the effects of the varying spot size and the energy chirp on the x-ray pulse duration, one without the energy chirp (i.e., a 67 MeV constant energy) and the other with a constant initial spot size $\sigma_\mathrm{r}=1.0~\nano\meter$, while other properties are consistent with the self-consistent beam. An optical undulator with $a_\mathrm{L}=1.8$ is used. As shown in Fig. \ref{fig:4}(d), a GW pulse is produced at the same location of the real beam case, which indicates that both the chirp and a varying $\sigma_\mathrm{r}$ can self-select which part of the beam lases. The pulse duration in both cases is approximately 1 as, which indicates that varying $\sigma_r$ and the chirp are essential for the zeptosecond pulse generation.

\begin{figure}
\includegraphics[width=\linewidth]{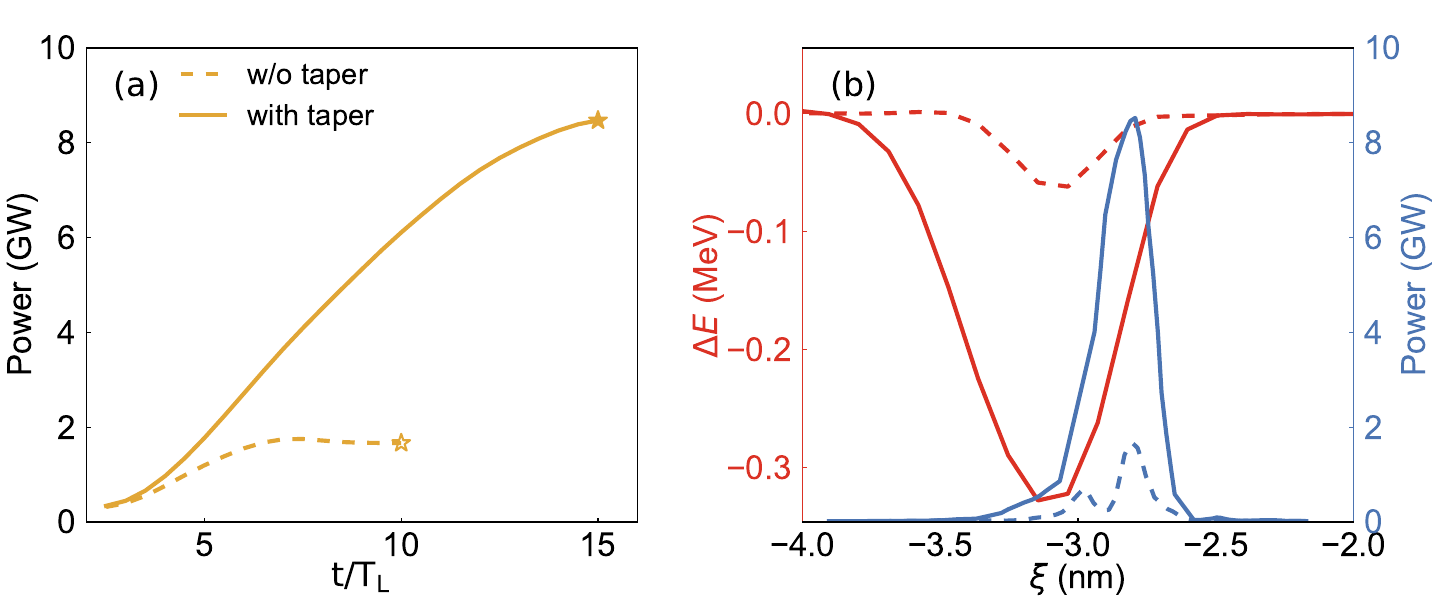}
\caption{\label{fig:5} Comparison between radiation pulses using a flattop laser (dashed lines) and a tapered laser undulator (solid lines). (a) The evolution of the radiation peak power. (b) The average energy loss per electron in different slices (red lines) and the radiation power distribution (blue lines) at $t=10T_\mathrm{L}$ and $t=15T_\mathrm{L}$ respectively.}
\end{figure}

Recent advances in manipulating the spatio-temporal laser pulse profiles could be utilized to improve the pulse intensity \cite{laser-shaping}. A tapered laser pulse whose $a_\mathrm{L}$ increases linearly from 1.8 to 2.8 in 15 periods is utilized to collide with the extreme beam. In contrast to the saturation of peak power after 7 periods in the flattop pulse case (dashed lines), the radiation peak power grows continuously up to 8.5 GW after 15 periods in the tapered case (solid lines), which is one order of magnitude higher, as shown in Fig. \ref{fig:5}(a). In the flattop case, because the resonant energy is fixed, when the radiation pulse slips out of the resonant region, the peak power stops growing. On the other hand, the resonant energy gradually increases in the tapered case, making the resonant region shift forward synchronously with the radiation, which enables the radiation to continuously extract energy from fresh electrons \cite{undulator-tapering,experimental-tapering}. As shown in Fig. \ref{fig:5}(b), in contrast to the flattop case, electrons in a substantially wide region lose $\lesssim0.3$ MeV energies in the tapered case and produce a high-power zeptosecond pulse. The radiation pulse has a FWHM duration of $700~\zepto\second$ and a spot size of 12.3 nm, with a pulse energy of $6.5~\nano\joule$, containing $10^6$ photons. 

The proposed scheme also has the potential to produce shorter pulses. The self-similar property of the PBA suggests brighter electron beams with higher density, shorter prebunched period, larger energy chirp and shorter spot size period can be produced if operating in higher solid-density plasmas. This should in turn produce shorter radiation pulses with similar power. Shorter and narrower e-beams but at less energy can be produced by increasing the plasma density in the laser-driven PBA. Recently proposed flying focusing techniques might enable accelerating electrons to $>$ GeV even in $>10^{19}~\centi\meter^{-3}$ plasmas \cite{dephasingless,phase-locked}. For example, if we improve $n_\mathrm{p1}$ and $n_\mathrm{p2}$ by a factor of 16, the 3D sizes of the GeV driver can be decreased by a factor of 4 and it thus can excite a nonlinear wakefield in a solid-density plasma with $n_\mathrm{p3}=9.6\times 10^{23}~\centi\meter^{-3}$ and finally produce high power $\sim100$ zs pulses with $\sim100$ keV photon energy.

In conclusion, we present a comprehensive and feasible way to produce an isolated intense zeptosecond pulse using an extremely dense and bright prebunched electron beam generated from staged PBAs. Both the spatial distribution and the energy chirp of the beam facilitate the generation of such ultrashort pulses. PIC simulations show that a hard x-ray pulse with a 700 zs duration and a 8.5 GW peak power can be generated in a tapered laser pulse. We emphasize that, despite the experimental complexity, our scheme can be implemented using currently achievable 100 TW laser systems. Such intense zeptosecond pulses can provide an essential tool for probing the dynamics of nuclear physics and QED processes and open a wide range of applications and configurations for ultrafast science. 

\begin{acknowledgments}
This work was supported by the National Grand Instrument Project (No. 2019YFF01014400), the National Natural Science Foundation of China (NSFC) (No. 12375147), Guangdong Provincial Science and Technology Plan Project (2021B0909050006), the Fundamental Research Funds for the Central Universities, Peking University, the CAS Project for Young Scientists in Basic Research (YSBR-115), and the U.S. Department of Energy under Contracts No. DESC0010064. C.J. supported by DOE grant DE-SC0010064:0011 and by UCLA. The simulations were supported by the High-performance Computing Platform of Peking University and the resources of the National Energy Research Scientific Computing Center.
\end{acknowledgments}




\begin{thebibliography}{62}
\bibitem [{EMc()}]{EMconstants}%
  \BibitemOpen
  \href@noop {} {}\bibinfo {note} {$e$ and $m_e$ are the electron charge and mass, $\epsilon_0$ is the vacuum permittivity, and $c$ is the speed of light in vacuum.}\BibitemShut {Stop}%
\bibitem [{ful()}]{fullexpressionofrho}%
  \BibitemOpen
  \href@noop {} {}\bibinfo {note} {$\rho\equiv\left(\frac{e^2 K^2[\mathrm{JJ}]^2n_\mathrm{b}}{32\epsilon_0\gamma^3mc^2 k_\mathrm{u}^2}\right)^{1/3}$, where $K$ and $k_\mathrm{u}$ are the normalized vector potential and the wavenumber of the undulator, and $[\mathrm{JJ}]$ is the coupling factor between the electrons and the radiation field and its value depends on $K$.}\BibitemShut {Stop}%
\end{thebibliography}
%

\end{document}